\documentclass[superscriptaddress,twocolumn,prl,floatfix,showpacs]{revtex4-1}

\usepackage{amsmath,bm,amssymb,amsfonts,graphicx,psfrag}
\usepackage{ulem,color}

\begin{document}

\title{Bending dynamics of semi-flexible macromolecules in 
isotropic turbulence}

\author{Aamir Ali}
\affiliation{Universit\'e Nice Sophia Antipolis, CNRS, 
Laboratoire Jean Alexandre Dieudonn\'e, UMR 7351,
06100 Nice, France}
\affiliation{Department of Mathematics,
COMSATS Intitute of Information Technology, Islamabad, Pakistan}
\author{Samriddhi Sankar Ray}
\affiliation{International Center for Theoretical Sciences, Tata Institute of Fundamental Research, Bangalore 560012, India}

\author{Dario Vincenzi}
\affiliation{Universit\'e Nice Sophia Antipolis, CNRS, 
Laboratoire Jean Alexandre Dieudonn\'e, UMR 7351,
06100 Nice, France}

\date{\today}

\begin{abstract}
We study the Lagrangian dynamics of semi-flexible macromolecules
in laminar as well as in homogeneous and isotropic turbulent
flows by means of analytically solvable stochastic models
and direct numerical simulations.
The statistics of the bending angle is qualitatively 
different in laminar and turbulent flows and exhibits a strong dependence
on the topology of the velocity field. 
In particular, in two-dimensional turbulence,
particles are either found in a fully extended or in a fully folded
configuration; in three dimensions, the predominant configuration is the
fully extended one.
\end{abstract}


\maketitle

The study of hydrodynamic turbulence and of turbulent transport has received
considerable impulse from the development of experimental, theoretical, and
numerical Lagrangian techniques~\cite{FGV01,TB09,PPR09,SC09}.  
The translational
dynamics of tracer and inertial particles is indeed related to the mixing
properties of turbulent flows~\cite{BBBCMT06} 
and has applications in geophysics
(atmospheric pollution, rain formation)~\cite{S03}, astrophysics (dynamo
effect, planet formation) \cite{CG95,A10}, and chemical
engineering~\cite{FT11}.  Over the last decade, attention has extended to the
Lagrangian dynamics of particles possessing additional degrees of freedom, such
as elastic polymers and nonspherical solid particles (see, e.g.,
Refs.~\cite{MV11,V13} and references therein).  In particular, the examination
of the extensional dynamics of polymers has provided information on the
coil-stretch transition and on turbulent drag reduction \cite{S09}; recent
studies of the dynamics of non-spherical particles in isotropic turbulence have
investigated the alignment and rotation statistics of such particles
\cite{PW11,PCTV12,GVP13,CM13,GEM14}.  In this Letter, we consider particles that
possess yet another degree of freedom, namely semi-flexible macromolecules 
that can bend
under the action of a non-uniform velocity field.  We model a semi-flexible
macromolecule as a trumbbell, i.e., as three beads connected by two rigid links
\cite{H74}; the stiffness of the trumbbell is controlled by an elastic hinge,
which prevents the particle from bending.  The trumbbell model (also known as
trimer or three-bead-two-rod model) qualitatively captures the viscoelastic
properties of suspensions of segmentally flexible macromolecules
\cite{H74,BHAC77,RZ84,R84,NY85,R87,GT94}.  This model also plays an important
role in statistical mechanics as the prototypical system for
showing that the infinite-stiffness limit of elastic bonds is singular
\cite{H74,GB76,H79,R79,H94}.  Moreover, an active variant of the trumbbell
model, the trumbbell swimmer, describes the swimming motion of
certain biological
microorganisms~\cite{NG04+GA08,NRY13} (see also Ref.~\cite{SN11}).  
We study the bending dynamics of trumbbells, in both laminar and fully 
turbulent flows, 
via analytical calculations and detailed numerical simulations, and
show that the statistics of the bending angle  
depends strongly on the topology of the flow.
In particular, in a two-dimensional
(2D) homogeneous and isotropic, incompressible
turbulent flow, the configurations in which the rods are parallel or
antiparallel are the most probable ones; as the amplitude of the velocity
gradient increases, these probabilities become sharper,
with the parallel configuration
becoming the most likely for very strong turbulence. By contrast,
in three dimensions, the antiparallel configuration 
always is the most probable one.

\begin{figure}[t]
\includegraphics[width=0.98\columnwidth]{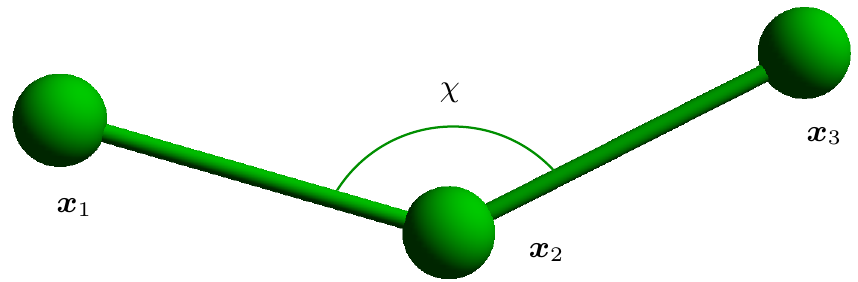}
\caption{(color online) The trumbbell model.}
\label{fig:angles}
\end{figure}

A trumbbell consists of three spherical beads joined by two 
inertialess rigid rods of equal length~$\ell$ immersed in a Newtonian 
fluid (Fig.~\ref{fig:angles}) \cite{H74,BHAC77}.
The drag force acting on each of the beads is given by the
Stokes law with a drag coefficient~$\zeta$.
An elastic hinge at the middle of the trumbbell models
the action of the entropic forces
which prevent the trumbbell from bending;
the force exerted by the elastic hinge is described by a harmonic potential.
The maximum end-to-end extension of the trumbbell, $2\ell$,
is sufficiently small for (a) it to experience Brownian fluctuations and
(b) the velocity gradient $\nabla\bm u$ to be
spatially uniform across the trumbbell.
The inertia of the beads and the hydrodynamic
interactions between them are disregarded. Furthermore,
the suspension is assumed to be sufficiently dilute; 
hence, particle-particle hydrodynamic interactions are negligible.

The positions of the beads are denoted as $\bm x_i$, $i=1,2,3$.
Under the above assumptions, the
position of the center of mass, $\bm x_c\equiv
(\bm x_1+\bm x_2+\bm x_3)/3$, evolves like that of a tracer,
i.e., $\dot{\bm x}_c(t)=\bm u(\bm x_c(t),t)$.
The separation vectors between the beads and the center of mass are defined as
$\bm r_\nu\equiv\bm x_\nu-\bm x_c$ with $\nu=1,2,3$.
However, the configuration of the trumbbell
in the reference frame of the center of mass
is more conveniently described in terms of a reduced set of $2(d-1)$ 
angular
coordinates $\bm q$, where $d$ is the dimension of the fluid~\cite{BHAC77}. 
For $d=2$, $\bm q=(\theta,\chi)$
where $0\leqslant\theta<2\pi$ 
gives the orientation of the vector $\bm x_3-\bm x_2$ with
respect to a fixed frame of reference and $0\leqslant \chi<\pi$ is the internal
angle between $\bm x_1-\bm x_2$ and $\bm x_3-\bm x_2$
(Fig.~\ref{fig:angles}).
By letting $\chi$ vary between 0 and $\pi$ only, we do not distinguish
between the two configurations that are obtained by exchanging
$\bm x_1$ and $\bm x_3$.
The separation vectors $\bm r_\nu$ are expressed in terms
of $\theta$ and $\chi$ as follows: 
\begin{equation}
\begin{array}{l}
\bm r_1=\frac{\ell}{3}\left(2\cos(\theta+\chi)-\cos(\theta),
2\sin(\theta+\chi)-\sin(\theta)\right),\\
\bm r_2=-\frac{\ell}{3}\left(\cos(\theta+\chi)+\cos(\theta),
\sin(\theta+\chi)+\sin(\theta)\right),
\\
\bm r_3=\frac{\ell}{3}\left(2\cos(\theta)-\cos(\theta+\chi),
2\sin(\theta)-\sin(\theta+\chi)\right).
\end{array}
\end{equation}
For $d=3$, $\bm q=(\alpha,\beta,\gamma,\chi)$ where 
$0\leqslant\chi<\pi$
is once again the internal angle between the two rods 
(Fig.~\ref{fig:angles}) and
$0\leqslant \alpha,\gamma< 2\pi$, $0\leqslant \beta<\pi$
are the Euler angles that specify the absolute orientation of the 
orthogonal triad $(\bm x_3-\bm x_1)\wedge(\bm x_2-\bm x_c)$,
$\bm x_3-\bm x_1$, $\bm x_2-\bm x_c$ with respect to a fixed coordinate
system (for the relation between $\bm r_\nu$ and $\bm q$ in three dimensions,
see Refs.~\cite{H74,BHAC77}).

The statistics of the configuration of the trumbell is specified by
the distribution function $\psi(\bm q;t)$, which is such that
$\psi(\bm q;t)Jd\bm q$ is the probability that, at time $t$,  the 
angular variables describing the configuration lie 
in the range $\bm q$ to $\bm q+d\bm q$. Here,
$J$ is the Jacobian of the
transformation from the coordinates $\bm r_\nu$ to the coordinates 
$\bm q$:
$J=1$ for $d=2$ and $J=\sin\chi\sin\beta$ for $d=3$. 
Furthermore, $\psi(\bm q;t)$ is normalized as $\int \psi(\bm q;t)Jd\bm q=1$.
For a deterministic flow,
$\psi(\bm q;t)$
satisfies the following convection-diffusion equation~\cite{BHAC77}
(summation over repeated indices is understood throughout this Letter):
\begin{multline}
\dfrac{\partial\psi}{\partial t}=-\dfrac{1}{J}
\dfrac{\partial}{\partial q^i}\left\{
J\mathcal{G}^{ij}
\left[\left(\kappa^{kl}(t)r_\nu^{l}\dfrac{\partial r_\nu^k}{\partial q^j}
-\dfrac{1}{\zeta}\dfrac{\partial\phi}{\partial q^j}
\right)\psi\right.\right.\\
-\left.\left.\dfrac{KT}{\zeta}\sqrt{h}
\dfrac{\partial}{\partial q^j}\left(\dfrac{J\psi}{\sqrt{h}}\right)\right]
\right\},
\label{eq:diffusion}
\end{multline}
where $\kappa^{ij}(t)=\partial^j u^i(t)$ is the velocity gradient
evaluated at $\bm x_c(t)$, 
$K$ is the Boltzmann constant, $T$ is temperature,
$\phi=\mu(\chi-\pi)^2/2$ is the bending potential,
$\mathcal{G}=\mathcal{H}^{-1}$ with
$\mathcal{H}^{ij}= \frac{\partial r^k_\nu}{\partial q^i} \frac{\partial r^k_\nu}{\partial q^j},$
and $h=\operatorname{det}(\mathcal{H})$. 
As $\bm q$ are angular variables,
$\psi(\bm q;t)$ satisfies periodic boundary conditions.

In the absence of flow ($\bm\kappa=0$) and of bending potential ($\mu=0$), the
stationary form of $\psi(\bm q;t)$ is $\psi_{\mathrm{st}}(\chi)=N\psi_0(\chi)$
with $\psi_0(\chi)=\sqrt{4-\cos^2\chi}$ \cite{H74} (throughout this Letter, $N$
denotes a normalization constant).
Thus, the most probable configuration is
that with the rods being perpendicular ($\chi=\pi/2$); the parallel ($\chi=0$)
and antiparallel ($\chi=\pi$) configurations are equally probable, but their
probability is about 15\% smaller.
The bending potential breaks the aforementioned symmetry. 
For $\mu\neq 0$, 
$\psi_{\mathrm{st}}(\chi)=N\psi_0(\chi)
\exp\left[-Z(\pi-\chi)^2/2\right]$ \cite{H74}, where $Z\equiv\mu/KT$ is a
stiffness parameter and determines the equilibrium configuration of the
trumbbell.  Thus, the probability of the $\chi=\pi$
configuration increases as $Z$ increases.  Finally, when the trumbbell is immersed in a
non-uniform flow ($\bm\kappa(t)\neq 0$) its dynamics results from the interplay
between the restoring action of the elastic hinge and the deformation by the
flow.

\begin{figure*}
\includegraphics[width=0.258\textwidth]{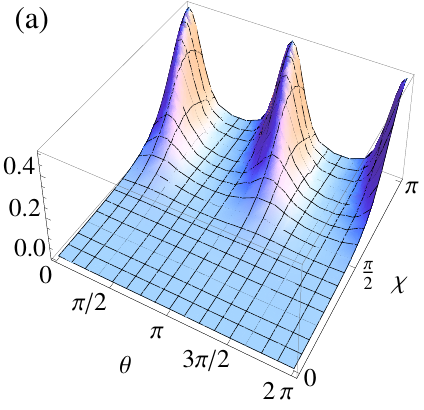}
\hfill
\includegraphics[width=0.25\textwidth]{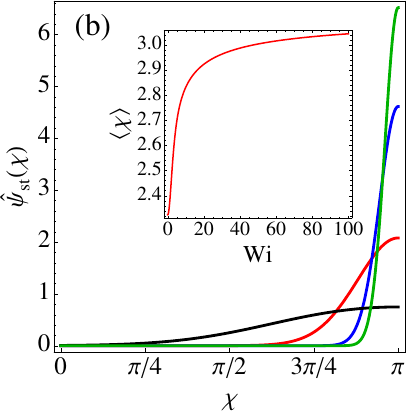}
\hfill
\includegraphics[width=0.264\textwidth]{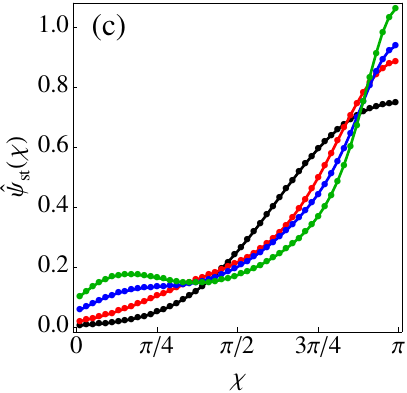}
\caption{(color online) 2D hyperbolic flow : (a)
surface plot of $\psi_{\mathrm{st}}(\theta,\chi)$ for $Z=1$ and $\mathrm{Wi}
=2$. For $Z=1$ and $\mathrm{Wi}=0$, 
$\langle\chi\rangle\approx 2.32$, slightly less than $3\pi/4$;
(b) Marginal distribution of $\chi$ for $Z=1$ and $\mathrm{Wi}=0$ (black), 
$\mathrm{Wi}=10$ (red), $\mathrm{Wi}=50$ (blue),
$\mathrm{Wi}=100$ (green).
The inset shows the average of $\chi$ vs. $\mathrm{Wi}$ for $Z=1$.
(c) Marginal distribution of~$\chi$ for $Z=1$ and $\mathrm{Wi}=0$ (black), 
$\mathrm{Wi}=10$ (red), $\mathrm{Wi}=50$ (blue), $\mathrm{Wi}=100$ (green) for a 2D simple shear flow. }
\label{fig:extensional}
\end{figure*}

We first examine the effect of pure stretching on the statistics
of the bending angle. Consider
a 2D hyperbolic velocity field $\bm u(x,y)=\sigma(x,-y)$. 
This flow is potential, i.e., $\bm\kappa=\bm\kappa^{\mathrm{T}}$.
Therefore, the stationary solution of Eq.~\eqref{eq:diffusion}
takes the form:
$\psi_{\mathrm{st}}(\theta,\chi)=N \sqrt{h}\exp[(\Phi-\phi)/KT]$,
where $\Phi=(\zeta/2)\sum_\nu\kappa^{ij}r^i_\nu r^j_\nu$ \cite{K44,H74}.
The function $\psi_{\mathrm{st}}$
is shown in Fig.~\ref{fig:extensional}(a)
for representative values of the parameters.
Since $\psi_{\mathrm{st}}$ 
depends on both $\chi$ and $\theta$, the statistics of the bending angle
varies according to the orientation of the trumbbell with respect to
the stretching direction of the flow. However, the
average effect of the hyperbolic flow can be understood by considering
the marginal distribution
$\widehat{\psi}_{\mathrm{st}}(\chi)\equiv
\int_0^{2\pi}\psi_{\mathrm{st}}(\theta,\chi)d\theta$, which can be written
explictly as follows:
\begin{equation}
\textstyle
\widehat{\psi}_{\mathrm{st}}(\chi)=N\psi_0(\chi)
\exp\left[-\frac{Z}{2}(\pi-\chi)^2\right]
I_0\left(\textstyle\frac{Z(1-2\cos\chi)\mathrm{Wi}}{3}\right),
\end{equation}
where $I_0$ is the modified Bessel function of the first kind of order 0 and
$\mathrm{Wi}=\sigma\tau$ (here $\tau=\zeta\ell^2/\mu$ 
is the characteristic time scale of the bending potential) \cite{foot1}. 
The Weissenberg number $\mathrm{Wi}$
measures the relative strength of the 
flow and of the bending potential. Since the function $I_0(z)$ is even, decreases with $z$ for $z<0$ and increases for $z>0$, 
on average the hyperbolic flow favors the $\chi=\pi$ configuration.
Hence it strengthens the action of the entropic forces
and this effect becomes stronger as $\mathrm{Wi}$ increases
(Fig.~\ref{fig:extensional}(b)).
An analogous analysis for a three-dimensional
(3D) incompressible hyperbolic flow shows that 
the stationary statistics of $\chi$ exhibits similar properties in
two and in three dimensions. 

In a hyperbolic flow, a trumbbell is subject only to pure stretching. However,
most practical flows have a rotational component as well.
A purely rotational planar flow, $\bm u(x,y)=\sigma(y,-x)$, has no effect 
on the stationary statistics of $\chi$. Indeed,
for this flow,
the convective term in Eq.~\eqref{eq:diffusion} has zero $\chi$-component,
and hence $\psi(\theta,\chi;t)$ converges in time to the stationary solution
of the $\mathrm{Wi}=0$ case.
However, the combined effect of stretching and rotation can influence
the bending dynamics. 
To illustrate this fact, we consider a simple
shear $\bm u(x,y)=\sigma(y,0)$, which is the superposition 
of a purely extensional component and of a purely rotational component
with equal magnitudes.
If the flow is not potential, no general 
analytical expression is available for $\psi_{\mathrm{st}}(\theta,\chi)$.
Thus we compute
$\widehat{\psi}_{\mathrm{st}}(\chi)$ from a Monte Carlo numerical 
simulation of Eq.~\eqref{eq:diffusion}. Figure~\ref{fig:extensional}(c) shows that
the $\chi=\pi$ configuration is the most probable one for all the values of $\mathrm{Wi}$ considered.
As $\mathrm{Wi}$ increases, the interplay
between rotation and stretching increases the probability of the small-$\chi$
configurations, unlike the purely extensional flow where it is negligible.
However, for a given $\mathrm{Wi}$,
the ability of the shear flow to bend a trumbbell is less than 
that of an hyperbolic flow with comparable $\mathrm{Wi}$.

In a random or turbulent flow, the extensional and rotational components
of the velocity gradient fluctuate in time and in space.
The above analysis suggests that the combination of these two components
may generate a complex bending dynamics.
In order to examine this point analytically,
we consider the Kraichnan random
flow~\cite{K68} in the Batchelor regime (see also Ref.~\cite{FGV01}).
The velocity gradient $\bm\kappa(t)$ is a delta-correlated-in-time
$(d\times d)$-dimensional
Gaussian stochastic process
with zero mean and correlation:
$\langle\kappa^{ij}(t)\kappa^{kl}(t')\rangle=
\mathcal{K}^{ijkl}\delta(t-t')$,
where $\mathcal{K}^{ijkl}=2\lambda[(d+1)\delta^{ik}\delta^{jl}-\delta^{ij}
\delta^{kl}-\delta^{il}\delta^{jk}]/[d(d-1)]$ and
$ \lambda$ is the maximum Lyapunov exponent of the flow.
The form of the tensor $\mathcal{K}$ ensures that the flow
is incompressible and statistically isotropic.
Although the assumption of temporal decorrelation is unrealistic,
the study of this flow has contributed in a substantial way
to the understanding of 
turbulent transport  (e.g., of passive scalars,  magnetic fields, and elastic polymers),
because it allows an analytical approach to this problem~\cite{FGV01}. 
We denote by $P(\bm q;t)$ the distribution
of the configuration of the trumbbell
with respect to the realizations of both $\bm\kappa(t)$ and thermal noise;
$P(\bm q;t)$ is normalized as $\int P(\bm q;t)Jd\bm q=1$.
A Gaussian integration by part of Eq.~\eqref{eq:diffusion}
(e.g., Ref.~\cite{F95}) yields the following equation
for $P(\bm q;t)$:
\begin{multline}
\dfrac{\partial P}{\partial t}=
\dfrac{\partial}{\partial q^i}\left\{
\dfrac{1}{2}\mathcal{K}^{klmn}\mathcal{G}^{ij}r_\nu^l\dfrac{\partial r_\nu^k}{\partial q^j}
\dfrac{\partial}{\partial q^a}\left(
\mathcal{G}^{ab}r_\rho^n\dfrac{\partial r_\rho^m}{\partial q^b}P\right)+
\right.\\
\left.\frac{\mathcal{G}^{ij}}{\zeta}
\left[\dfrac{\partial\phi}{\partial q^j}P
+KT\sqrt{h}
\dfrac{\partial}{\partial q^j}\left(\dfrac{P}{\sqrt{h}}\right)\right]\right\},
\label{eq:FPE-general}
\end{multline}
which must be solved with periodic boundary conditions.

\begin{figure}
\includegraphics[width=0.5\columnwidth]{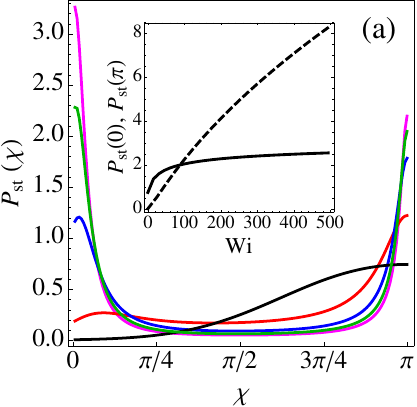}%
\hfill%
\includegraphics[width=0.49\columnwidth]{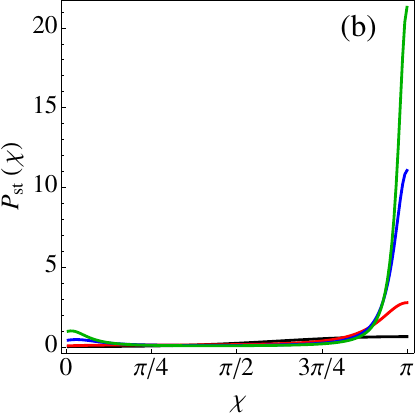}
\caption{(color online) Stationary distribution function
of~$\chi$ for the Batchelor-Kraichnan flow 
with $Z=1$ and $\mathrm{Wi}=0$ (black), 
$\mathrm{Wi}=10$ (red), $\mathrm{Wi}=50$ (blue), $\mathrm{Wi}=100$ (green),
and $\mathrm{Wi}=150$ (magenta) in (a) two and (b) three dimensions. 
The inset in (a) shows the values of $P_{\mathrm{st}}(\chi)$ at $\chi=0$ (dashed line)
and $\chi=\pi$ (solid line) as a function of $\mathrm{Wi}$.} 
\label{fig:random}
\end{figure}

Let us first consider the 2D case.
When $\bm q=(\theta,\chi)$, $J=1$,  
and the corresponding $\mathcal{G}$ are substituted
into Eq.~\eqref{eq:FPE-general}, we can  rewrite Eq.~\eqref{eq:FPE-general}
as a Fokker-Planck equation (FPE)
in the two variables $\theta$ and $\chi$. 
The drift and diffusion coefficients, however, do not depend on $\theta$ as a consequence of the
statistical isotropy of the flow.
Hence the stationary distribution of the configuration is a function of
$\chi$ alone and is the stationary solution of the following 
FPE in one variable:
\begin{equation}
\partial_s P=-\partial_\chi (V_\chi  P)
+\textstyle\frac{1}{2}\partial^2_{\chi\chi} (D_{\chi\chi} P),
\label{eq:FPE-2D}
\end{equation}
where $s=t/\tau$,
\begin{multline}
V_\chi(\chi)=\frac{12\sin\chi}{Z\left(2-\cos\chi\right)\left(2+\cos\chi\right)^2}
+\frac{6\left(\pi-\chi\right)}{\left(2+\cos\chi\right)} +
\\
2\mathrm{Wi}\,
\frac{\sin\chi[5+\cos\chi-11\cos(2\chi)-\cos(3\chi)]}%
{\left(2-\cos\chi\right)\left(2+\cos\chi\right)^3},
\label{eq:drift}
\end{multline}
and
\begin{equation}
D_{\chi\chi}(\chi)=\frac{12}{Z\left(2+\cos\chi\right)}
+\mathrm{Wi}\,\frac{16\sin^2\chi}{\left(2+\cos\chi\right)^2}.
\label{eq:diff}
\end{equation}
In a random flow, the time scale associated with stretching is given by 
$\lambda$; accordingly, in Eqs.~\eqref{eq:drift} and \eqref{eq:diff}
$\mathrm{Wi}=\lambda\tau$.
The stationary solution of Eq.~\eqref{eq:FPE-2D} that satisfies periodic
boundary conditions is~\cite{R89}:
\begin{equation}
P_{\mathrm{st}}(\chi)=\dfrac{N}{D_{\chi\chi}(\chi)}\exp
\left(2\int_0^\chi \dfrac{V_\chi(\eta)}{D_{\chi\chi}(\eta)}\,d\eta\right).
\label{eq:pst}
\end{equation}
The function $P_{\mathrm{st}}(\chi)$ is plotted in Fig.~\ref{fig:random}(a)
for different values of $\mathrm{Wi}$. 
For small $\mathrm{Wi}$, the most probable configuration is the
antiparallel one, as in the hyperbolic flow. However,
as $\mathrm{Wi}$ increases, 
a second peak emerges near $\chi=0$, while intermediate values of
$\chi$ become less and less probable. At large Wi, 
$P_{\mathrm{st}}(\chi)$ consists of two narrow peaks, one at $\chi=\pi$
and the other approaching $\chi=0$, with the latter becoming more and more
pronounced as Wi increases (Fig.~\ref{fig:random}(a)). 

It is important to ask if the results above carry over to the 3D Batchelor-Kraichnan flow.
In three dimensions, Eq.~\eqref{eq:FPE-general} can be rewritten
as a FPE in the four variables
$\alpha,\beta,\gamma,\chi$.
As a consequence of the statistical isotropy of the flow, the stationary 
distribution of the configuration, 
$P_{\mathrm{st}}(\bm q)$, depends only on $\chi$.
In particular, $P_{\mathrm{st}}(\chi) \sin\chi$ 
is again the stationary solution of a FPE in one variable.
Figure~\ref{fig:random}(b) shows $P_{\mathrm{st}}(\chi)$ for different values
of $\mathrm{Wi}$. 
A small peak at 
$\chi=0$ emerges only for very large values of $\mathrm{Wi}$; 
the $\chi=\pi$ configuration prevails for all values of $\mathrm{Wi}$
(Fig.~\ref{fig:random}(b)).
Thus, 
the statistics of $\chi$ 
is different in two and three dimensions. In particular, 
for $d=3$, the behavior of the distribution of $\chi$ is
qualitatively similar to that in the purely extensional flow.

Are the results obtained so far for model flows valid in realistic, turbulent 
flows? To ensure that the qualitative properties of the statistics of $\chi$ do not
depend on the Gaussianity and temporal decorrelation of the
Batchelor-Kraichnan flow, we perform a Lagrangian direct numerical simulation of the
trumbbell model in a 2D, incompressible turbulent
flow.  Our direct numerical simulation of the forced (at large scale) 
incompressible, 2D Navier-Stokes equation
uses a pseudospectral method~\cite{pseudo} and a second-order, exponential 
Runge-Kutta scheme ~\cite{cox+mat02}, with $1024^2$ collocation points and the $2/3$ dealiasing
rule on a 2$\pi$-periodic square domain. 
We present representative results from a simulation where the choice of the forcing amplitude, 
the coefficient of the Ekman friction, and the kinematic viscosity yield 
a Taylor-microscale Reynolds number $\mathrm{Re}_{\lambda} \approx 827$, in the non-equilibrium, statistically steady state; 
we have done several other simulations 
with different $\mathrm{Re}_{\lambda}$  and forcing scales which yield results consistent with the ones presented here.
This flow is seeded with $10^3$ independently evolving trumbbells 
each of whose center of mass evolves as a Lagrangian tracer particle; 
the velocity at $\bm x_c$, is evaluated from the Eulerian velocity by using a
bilinear-interpolation scheme~\cite{pp11}.
The Lagrangian evolution of the configuration of the
trumbbell is once again computed from a Monte Carlo simulation of
Eq.~\eqref{eq:diffusion}.  We define 
the Weissenberg number as
$\mathrm{Wi}=\tau/\tau_{\mathrm{min}}$, where $\tau_{\mathrm{min}}$
is the smallest time scale associated with the direct cascade of enstrophy
and is estimated as the minimum value of $(\sqrt{k^3 E(k)})^{-1}$ \cite{BS02}
(here $E(k)$ denotes the isotropic kinetic-energy spectrum).  
\begin{figure}[t]
\includegraphics[width=0.48\columnwidth]{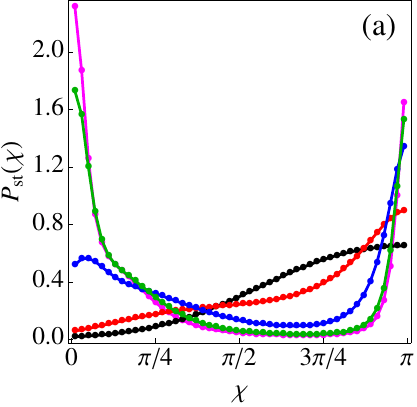}%
\hfill%
\includegraphics[width=0.518\columnwidth]{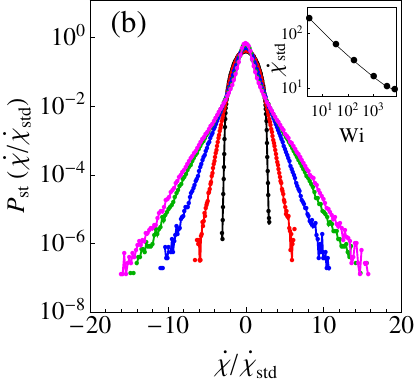}
\caption{(color online) 2D homogeneous, isotropic, turbulence: 
(a) distribution of $\chi$ for $Z=1$ and $\mathrm{Wi}\approx 3$ (black), 
$\mathrm{Wi}\approx 167$ (red), $\mathrm{Wi}\approx 10^3$ (blue), 
$\mathrm{Wi}\approx 3.3\times 10^3$ (green),
$\mathrm{Wi}\approx 6.6\times 10^3$ (magenta) and (b) distribution 
of $\dot \chi/\dot \chi_{\rm std}$ for $Z=1$ for different values of 
$\mathrm{Wi}$ [same as in panel (a)] and the dependence of $\dot \chi_{\rm std}$ on  $\mathrm{Wi}$.} 
\label{fig:turb}
\end{figure}
Figure~\ref{fig:turb}(a) shows the distribution 
function of the
internal angle $\chi$; the statistics of $\chi$ qualitatively agrees with that
calculated for the 2D Batchelor-Kraichnan random flow. In particular, we 
find that at small $\mathrm{Wi}$, the antiparallel 
configuration is favoured; with increasing $\mathrm{Wi}$, 
the probability of being in a parallel configuration increases significantly and eventually, 
at extremely large values of $\mathrm{Wi}$, the distribution function
displays strong peaks at both 
$\chi \approx 0$ and $\chi \approx \pi$.  We also calculate 
the distribution function 
of the rate of change of the angle  $\dot{\chi}$, normalised by its 
standard deviation  $\dot \chi_{\rm std}$, for various values of $\mathrm{Wi}$. 
With increasing $\mathrm{Wi}$ (Fig.~\ref{fig:turb}(b)), the distribution
function, which is symmetric 
and peaked at $\dot{\chi} \approx 0$, develops exponential tails. The standard deviation itself, $\dot \chi_{\rm std}$, 
is a decreasing function of  $\mathrm{Wi}$ (Fig.~\ref{fig:turb}(b), inset).


In this Letter, by using analytical calculations and numerical simulations, we
have shown how the conformation of semiflexible
macromolecules, modelled via the trumbbells,
can change with $\mathrm{Wi}$ as well as the flow topology.  
The statistics of the bending angle in turbulent or random flows
depends strongly on the space dimension and
is qualitatively different from that observed 
in simple laminar flows. This fact suggests that the rheology of
semiflexible macromolecules also exhibits a strong dependence on the
turbulent character of the flow and on its spatial dimension.
The recent advances in Lagrangian experimental techniques
allow extensive laboratory studies of the deformation of macromolecules in 
turbulent flows~\cite{TB09,S09}.
We hope our work would lead to further experiments directed towards the
study of the Lagrangian dynamics of semi-flexible macromolecules in turbulent
flows and
the non-Newtonian properties of turbulent suspensions of such particles.

\acknowledgments
The authors acknowledge the financial support of the EU COST Action MP0806
``Particles in Turbulence''.  The work of AA was supported by the Erasmus
Mundus Mobility with Asia (EMMA) programme.  SSR and DV acknowledge the support
of the Indo-French Center for Applied Mathematics (IFCAM).  SSR acknowledges
support from EADS Corporate Foundation Chair awarded to ICTS-TIFR and TIFR-CAM.
DV would like to thank the International Center for Theoretical Sciences of the 
Tata Institute of Fundamental Research (ICTS-TIFR)
and the Department of Physics of the Indian Institute of Science (IISc) in 
Bangalore for the kind hospitality.


\begin{thebibliography}{10}
\bibitem{foot1} 
For $\mu=0$, Eq.~\eqref{eq:diffusion} describes Hassager's free-hinge model~\cite{H74}; the
statistics of $\chi$ for the free-hinge model can be derived from our results
by letting $Z$ tend to zero while keeping $Z\mathrm{Wi}$ constant.
\end{thebibliography}
\end{document}